# CoreUI: Interactive Core Training System with 3D Human Shape

Haoran Xie[*]   Atsushi Watatani   Kazunori Miyata

Japan Advanced Institute of Science and Technology

xie@jaist.ac.jp

**Abstract**

We present an interactive core training system for core training using a monocular camera image as input in this paper. It is commonly expensive to capture human pose using depth cameras or multiple cameras with conventional approaches. To solve this issue, we employ the skinned multi-person linear model of human shape to recover the 3D human pose from 2D images using pose estimation and human mesh recovery approaches. In order to support the user in maintaining the correct postures from target poses in the training, we adopt 3D human shape estimation for both the target image and input camera video. We propose CoreUI, a user interface for providing visual guidance showing the differences among the estimated targets and current human shapes in core training, which are visualized by markers at ten body parts with color changes. From our user studies, the proposed core training system is effective and convenient compared with the conventional guidance of 2D skeletons.

## 1. Introduction

Human pose estimation from a single image has been explored intensively in the fields of computer vision and computer graphics. Particularly, the recent advances in deep learning approaches are helping to speed up research progress on this topic greatly such as OpenPose [1] and DensePose [2]. Benefiting from the sophisticated techniques of human pose estimation, a potential and useful application area is the training system where the pose recognition plays an important role. The existing training support systems using 3D sensing technology are too expensive to acquire by common users, such as the training system from Fujitsu [3]. Also, the existing training systems are mostly developed for professional sports training purposes, and the user interfaces are usually complex and difficult to use. To solve these issues, we aim to provide a user-friendly user interface for visual feedback in sports training using state-of-the-art pose estimation approaches. In contrast to 3D sensing techniques using laser sensors and depth cameras, our proposed system only requires video frames from the normal web camera.

In this work, we particularly focus on the support system for core training, a popular training style for the young generation. In core training, the human core plays a significant role in body movement and balance, and it is the central part of the human body. The training of human trunk muscles has been executed and emphasized by professional athletes, and it is being accepted by the general public nowadays. According to the fitness report from the *Worldwide Survey of Fitness Trends*, core training has been the most popular training style being practiced in the world since 2010 [4].

Core training is recognized as a practical and simple training method and can be completed indoor on an individual basis every day. The training method is effective in training body muscles around trunk parts. To maximize the training effect of core training, it is very important to maintain correct poses [5], and training in an incorrect pose can lead to injury. However, it is difficult for an individual user to achieve the correct pose in training. Therefore, our research motivation is to guide the user to achieve a target pose from the current pose estimated from a single camera image, as shown in Figure 1. To solve this issue, we propose a straightforward and simple user interface for use by an individual in performing core training.

In the training user interface, we adopt a skinned multi-person linear (SMPL) model of human shape to show the differences between target and current poses (Figure 1). In the previous work about the training support system, they usually utilized depth sensor or multiple cameras in capturing the human pose [6,7]. In contrast to the expensive devices used in previous systems, we estimate the human pose from a normal web camera using deep learning based approaches. Additionally, the existing training system mainly adopted the human skeleton for visual feedback. It is difficult to understand the differences between target and current poses with skeleton guidance because the human body is in volumetric shapes. Therefore, we present an SMPL-based pose estimation with 3D human meshes to solve this issue.

In this paper, we propose CoreUI, a training system to support core training for common users in daily usage [8]. First, we estimate the 3D human shape and pose from a single image using OpenPose and human mesh recovery approach, which can also be the frames from a normal web camera. Afterward, we present the user interface for loading target images and providing visual feedback to the user during the training process. As shown in Figure 1, we employ two viewpoints of human shapes to easily understand the pose differences.

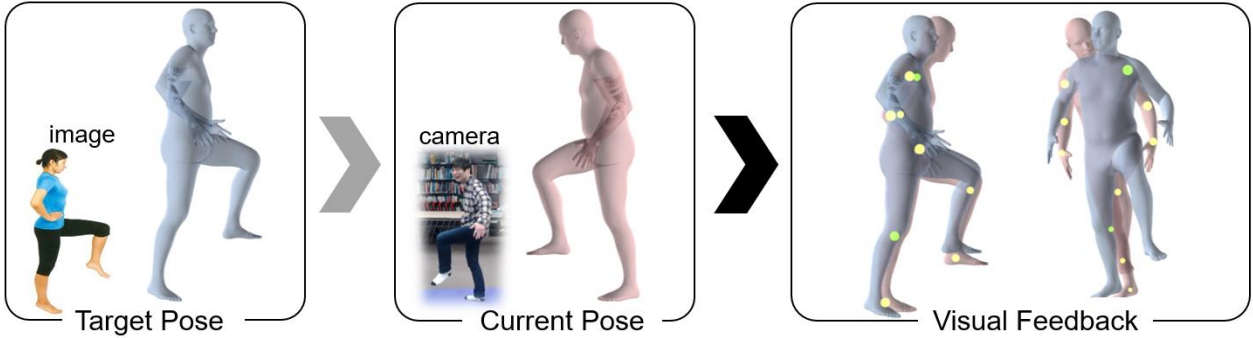

Figure 1. We propose a visual feedback system for core training which can guide the user in achieving the target pose from an input image. The system only requires a standard web camera to capture current pose in 3D and estimates the 3D human shape for further guidance.

The main contributions of this work are listed as follows: (1) we propose a training system for supporting core training for common user using a single web camera; (2) we provide an interactive user interface for carrying out training guidance on 3D human shapes and poses using deep learning approaches. We believe that the proposed system can also be suitable for other sports in an inexpensive and straightforward way.

## 2. Related Works

In this section, we briefly discuss the related works in the user interfaces for daily activity supports, human pose estimation, and sports training systems.

### 2.1 User Interface

The research work of user interface has been extensively explored in computer graphics and human interfaces. The conventional user interface is graphical user interface (GUI) with a long story from the birth of the computer, which is useful in supporting our creative activities, such as 2D painting [9] and 3D functional design [10]. Since the rapid progresses achieved in space sensing and wearable techniques, the user interfaces for supporting our daily activities have been proposed, such as selfie-taking [11] and golf training [12]. Besides the software developments, a supporting user interface can also be designed as augmented devices for amplifying the user abilities [13].

### 2.2 Human Pose Estimation

Human pose estimation is an important topic in computer vision. Recently, numerous deep learning based estimation approaches have been proposed. OpenPose can provide real-time 2D pose estimation using a nonparametric representation to learn the body parts in an image dataset [1]. DensePose applied fully convolutional neural networks and region-based models to map the RGB image to the 3D surface of the human body [2]. Human mesh recovery model was introduced for constructing 3D human meshes from a single image by minimizing the reprojection loss of key points in the image [14]. From video frames, deep pose estimation can be used to learn character dynamics with deep reinforcement learning [15]. Human mesh recovery model can also be used to estimate continuous 3D human shape from video clips in a semi-supervised manner [16]. In this work, we employ these state-of-the-art approaches for core training support and provide a user-friendly user interface for visual guidance.

### 2.3 Sports Training System

The common training systems are using depth sensors for pose estimate. Kinect was used to acquire depth information and color image simultaneously for detecting user's training behaviors in the proposed training system of abdominal muscle exercises [6]. The target pose is represented using a skeleton image, and the estimated calorie consumption is displayed. A similar training system was proposed for hurdle practice for judging the appropriate motion and correct pose during the training process [7]. A yoga self-training system was proposed for instructing the user to perform correct poses using Kinect [17]. A similar training system has been developed for gym fitness using real-time guidance and action assessment [18]. The training system of archery practice was proposed for form improvement in a head-mounted display with two RGB cameras [19]. A computer vision based yoga training system was developed by analyzing user pose from both front and side views using two cameras and the system can be used to display the body contour, skeleton, dominant axes, and feature points information [20]. Existing training systems usually require a depth sensor or multiple cameras with careful registration. In this study, we employ a web camera as the system input, which is inexpensive and ubiquitous in our daily life.

Virtual reality (VR) techniques have been used for sports training with feasible haptic interactions [21]. A training system for ball sports was presented with accurate physical models, real-time simulations and user interface training using both software and hardware components [22]. Another similar work introduced the table tennis simulation using projection system and tracking system [23]. A recent study developed the virtual environment to train rugby ball passing skills [24]. The training systems in VR aim to produce similar training

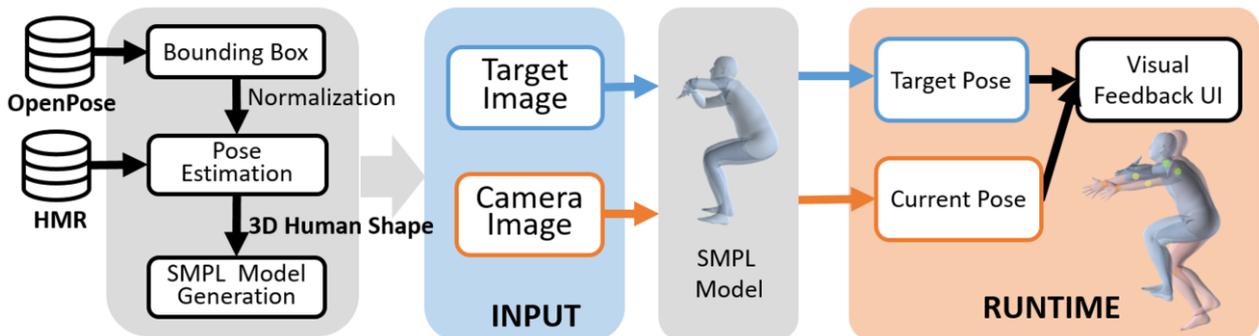

Figure 2. Framework of the proposed core training system. The system input is the user-provided target image and the camera image in each frame. In the runtime, the user interface for visual feedback used in the training process is designed to provide guidance to the user and to prevent injury that could result from incorrect poses.

experience as in normal real-life, however, the training guidance to a target pose was absent.

## 3. System Overview

In this section, we describe the whole framework of our core training system, and the proposed user interface provides visual guidance to the user for achieving the correct target poses.

### 3.1 System Framework

The framework of the proposed core training system consists of two main components: pose estimation and visual feedback (Figure 2).

For pose estimation, we employ the OpenPose method to obtain the bounding box of the human region in the input image and human mesh recovery approach to generate 3D human shape from the input. With the correction by the bounding box after the normalization process, a human mesh based on SMPL model can be generated from a single input image.

For visual feedback, a runtime user interface is proposed to guide the user on the differences between target pose and the current pose from the captured frame. Both target and current poses are estimated from a single image using the pose estimation method discussed above.

### 3.2 User Interface

Figure 3 shows our proposed user interface, CoreUI, an interactive training system for performing core training. The target image and the camera image are located in the middle part of the interface. In the two view windows of CoreUI, the 3D human models can be generated from camera images using pose estimation in the predefined viewpoints. The 3D human model in blue color denotes the target 3D pose while the 3D human model in red color represents the current pose from the camera image. In the user interface of CoreUI, the current pose model and the target pose model are displayed in a superimposed manner. Different colors of markers are used to indicate the differences between current and target models using for training guidance. We used red, orange, yellow and green-yellow colors in our implementation to represent distances ranging from far to close.

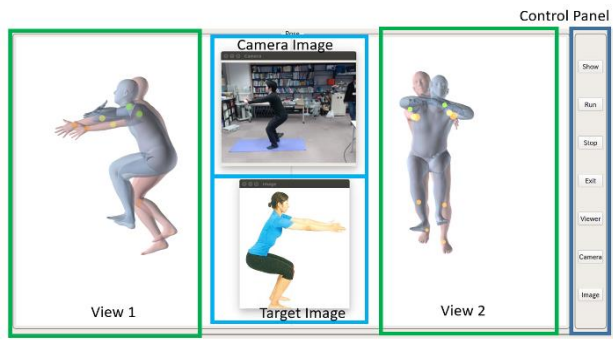

Figure 3. Proposed user interface for core training.

## 4. 3D Shape and Pose Estimation

To obtain 3D human shape and pose from a single image, the proposed system adopts pose estimation based on OpenPose to realize the bounding box of the human region. It also employs human mesh recovery to estimate the 3D human pose. For the generation of 3D human mesh, an SMPL model is used.

### 4.1 SMPL Model

OpenPose can be used to detect the multiple persons' pose from an input image using a multi-stage convolutional neural network (CNN). SMPL model can represent various human postures and shapes [25], which are defined by the triangular meshes $M(\theta, \beta)$ in $R^{3 \times N}$. $N = 6,980$ denotes the number of mesh vertices. In an SMPL model, human pose $\theta$ is defined by the skeleton rig with $K = 23$ joints and $\theta$ in $R^{3 \times K+3}$. Therefore, pose $\theta$ has $3 \times K + 3 = 72$ parameters including 3 for each part and 3 for root orientation. $\beta$ in $R^{10}$ denotes shape parameters for 3D human mesh model in 10 dimensions of shape space using principal component analysis. Figure 4 presents the 3D meshes and SMPL models in different $\theta$ and $\beta$ values.

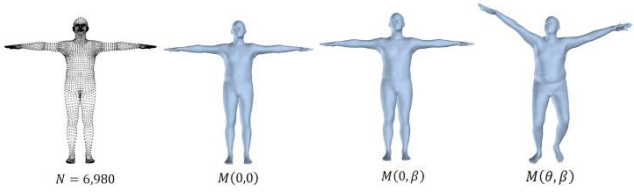

Figure 4. SMPL model of human mesh in different poses.

Human mesh recovery (HMR) model is a deep learning based approach for reconstructing 3D human model from a single RGB image. The human mesh model SMPL can be generated from 9 DoF camera pose (translation, rotation, and scaling) and SMPL parameters pose $\theta$ and shape $\beta$, which can be evaluated using a 3D regression module. In the discriminator network, the natural human shapes are distinguished if unnatural joint bending and unnatural body shapes such as too thin exist.

### 4.2 Normalization

In this study, we define an image normalization step before conducting pose estimation on the image. The reason for performing image normalization is to ensure that the learning dataset of HMR used in pose estimation is unified under the following two conditions: diagonal size of the bounding box is around 150 pixels; image size is 224×224 pixels. Because the bounding box of the human region in the input image is necessary for the implementation, we adopt OpenPose for generating the bounding box using 3D positions of all joints. Finally, we applied HMR model for pose estimation on the normalized images.

## 5. Visual Feedback

In this section, we discuss user interface for human shape generation and the training guidance of the visual feedback used for core training.

### 5.1 Human Shape Generation

In the proposed system, the user can predefine the easy-to-see viewpoints on the pre-training user interface, as shown in Figure 5. The user can first select the target image with a standard pose to be used for core training. Afterward, the system generates a 3D human model in two view windows using the proposed pose estimation approach. For each view window, the user can select the good viewpoints in his/her judgment. Finally, the system saves the predefined set of viewpoints.

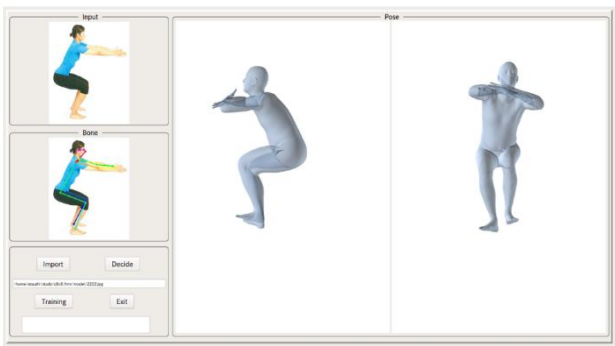

Figure 5. User interface for target shape generation.

### 5.2 Training Guidance

Herein, we present details of the training guidance implementation for the proposed training user interface. To display the target and current 3D human models in a superimposed manner, we rendered two models simultaneously using OpenGL. In the implementation, we simply aligned target and current models by coinciding with the origins of two 3D human models at the waist position. Although a more accurate alignment algorithm such as iterative closest point method may be helpful, we considered the simplicity of the proposed approach as a trade-off between accuracy and computation cost.

For representing the differences between target and current 3D models, we adopt color visualization on predefined 10 marker positions on the human body. We set the marker positions at both hands and elbows, shoulders, knees, and ankles because the differences between the two models are more apparent at the end parts than the human trunk. We did not choose hip joints on the body because the differences in hip points are not apparent due to coincided waist positions. We used shoulders, elbows, and hands positions for a good representation of the upper body postures, which are the most important parts in core training.

To obtain the marker positions using SMPL-based human model, we superimposed the aforementioned joint positions with SMPL mesh model to determine the start and end points for each marker, as shown in Figure 6.

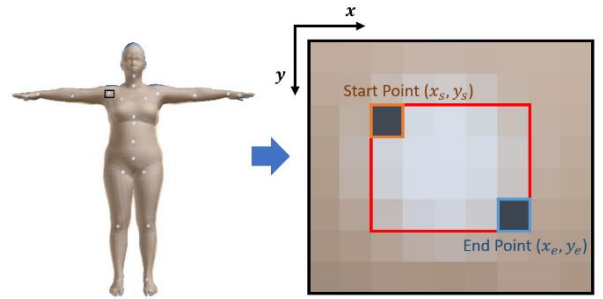

Figure 6. Start and end points in the image coordinate.

To obtain the mesh vertices of markers, we substituted the transformation matrix of perspective projection from the 3D model to the output image. The relation existing in the perspective projection can be expressed as follows.

$$\begin{pmatrix} x_{image} \\ y_{image} \\ 1 \end{pmatrix} = \begin{pmatrix} \frac{fX_{model}}{Z_{model}+2} + C_x \\ \frac{fY_{model}}{Z_{model}+2} + C_y \\ 1 \end{pmatrix}$$

where $f = 500$ denotes the focal length, $C_x = 332.50$ and $C_y = 325.00$ denote the optical center in x and y directions, respectively. $x_{image}$ and $y_{image}$ denotes the pixel positions in the output image and ($X_{model}$, $Y_{model}$, $Z_{model}$) represents the 3D vertex positions on human model. We search the vertex positions for all markers using the following constraints.

$$\begin{cases} X_{model} > \dfrac{x_s - C_x}{f}(Z_{model} + 2) \\ X_{model} < \dfrac{x_e - C_x}{f}(Z_{model} + 2) \\ Y_{model} > \dfrac{y_s - C_y}{f}(Z_{model} + 2) \\ Y_{model} > \dfrac{y_e - C_y}{f}(Z_{model} + 2) \end{cases}$$

Where $x_s$ and $y_s$ denotes the start point position, $x_e$ and $y_e$ denotes the end point position in the image coordinate as shown in Figure 6. Figure 7 shows the mesh vertex positions of all markers on SMPL human mesh used in the implementation.

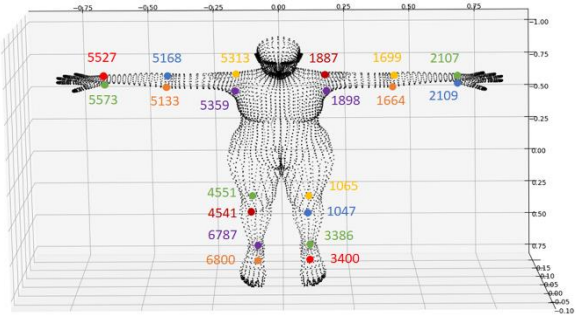

Figure 7. Mesh vertices for each marker position.

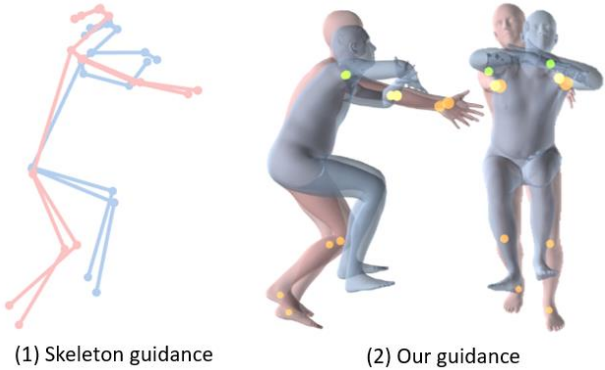

(1) Skeleton guidance     (2) Our guidance

Figure 8. Two types of training guidance used in the user study with skeleton guidance (1) and the proposed guidance approach (2).

Finally, we calculate the difference, $d_e$ in vertex positions with four colors. The color of marker point is set to be as follows.

$$\begin{cases} red\ if\ 0.5 \leq d_e < \infty; \\ orange\ if\ 0.25 \leq d_e < 0.5; \\ yellow\ if\ 0.1 \leq d_e < 0.25; \\ green\ yellow\ if\ 0 \leq d_e < 0.1 \end{cases}$$

## 6. User Study

To verify the effectiveness and feasibility of the proposed visual feedback system, we compared the proposed system with the conventional visual feedback of skeleton information, as shown in Figure 8. The skeleton guidance is usually employed in training support system with depth sensor such as Microsoft Kinect [6,7,17,18]. Because of the camera view of a depth sensor, only one viewpoint is available in skeleton guidance. However, the proposed guidance approach provides two viewpoints, thus, benefiting from the use of 3D human shape.

To avoid the effects of different hardware, we used the proposed pose estimation method to obtain skeleton information in the user study. In the skeleton guidance, we adopted 18 joints obtained from COCO dataset used in OpenPose [26]. They include ears, eyes, nose, neck, shoulders, elbows, hands, hips, knees and ankles.

### 6.1 Training Tasks

We had eight participants in the user study, all of whom are male graduate students around 25 years old. The participants were asked to do core training with the two visual guidance methods.

In the user study, we selected four core training tasks in different target poses including lateral pose, supine pose, and two standing poses as common training tasks [27]. Figure 9 shows the target pose image in 2D, the skeleton guidance obtained using the proposed pose estimation method, and the 3D human shapes and poses in two viewpoints obtained using the proposed visual feedback system. Note that the human arms of human shapes happened to be curved due to the limitation of the state-of-the-art shape estimation algorithm.

We asked the participants to complete the core training tasks for four times with different combinations of training guidance systems and tasks. The task combinations are presented in Table 2. It corresponds to two training guidance systems, as shown in Figure 8 and four assigned core training tasks, as shown in Figure 9.

|  | A | B | D | C | E | F | G | H |
|---|---|---|---|---|---|---|---|---|
| 1st time | 2-1 | 2-3 | 1-4 | 1-2 | 2-1 | 2-3 | 1-4 | 1-2 |
| 2nd time | 2-2 | 2-4 | 1-3 | 1-1 | 2-2 | 2-4 | 1-3 | 1-1 |
| 3rd time | 1-3 | 1-1 | 2-2 | 2-4 | 1-3 | 1-1 | 2-2 | 2-4 |
| 4th time | 1-4 | 1-2 | 2-1 | 2-3 | 1-4 | 1-2 | 2-1 | 2-3 |

Table 1. The training guidance and tasks for each participant for four times. For instance, the participant C at the third time (2-4) was asked to complete core training task 4 (Figure 9) under the guide 2 of the proposed guidance (Figure 8).

### 6.2 Training Evaluation

To make a quantitative evaluation of the two training guidance systems, we define two metrics of training accuracy and time. To account for training accuracy, we adopt root mean square error ($RMSE$) to evaluate the differences in the target and current poses. We obtained the following formulation of $RMSE_i$. at time $t_i$.

$$RMSE_i = \sqrt{\frac{1}{N}\sum_{k=1}^{N}(\vec{E}_k(t_i) - \vec{M}_k(t_i))}$$

where N = 5,628, denotes the number of model mesh vertices except the head part. Because the user performs pose correction while looking at the display screen, we excluded the mesh vertices of the head part in $RMSE$ calculation due to the possibility of inappropriate head movement. $E_k$ and $M_k$ are k-th vertex positions on current and target human models at time $t_i$. We define the training accuracy R as follows.

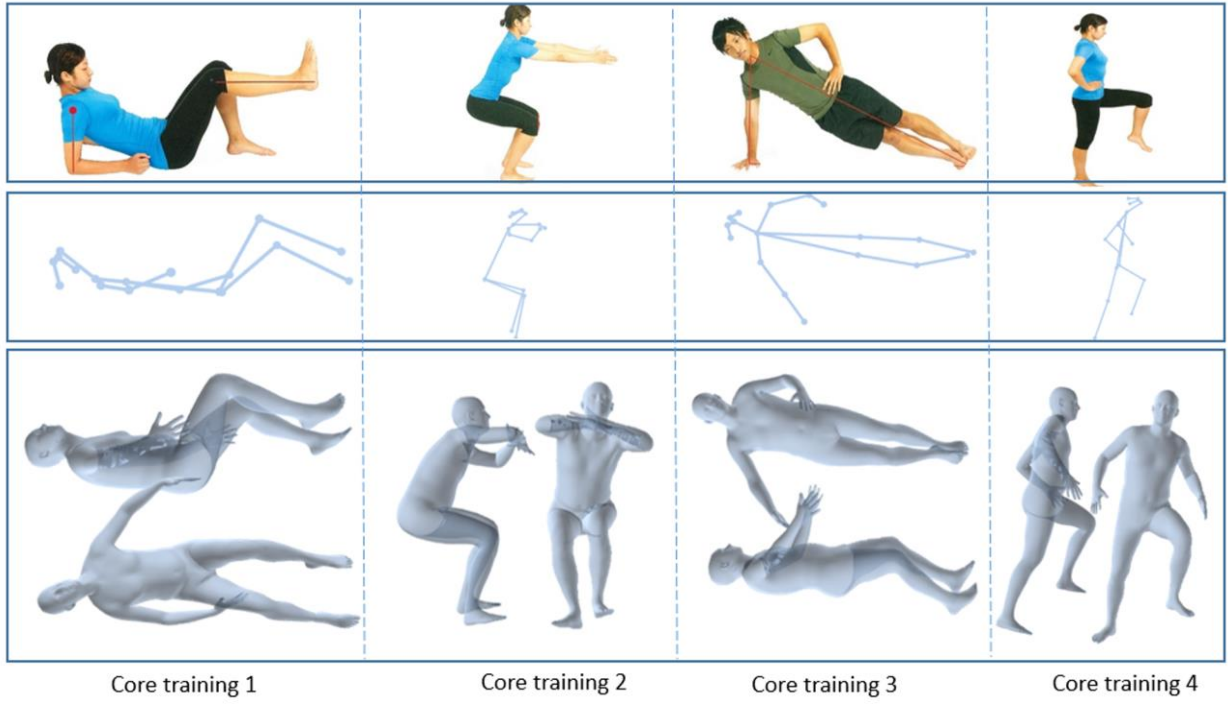

Figure 9. Core training tasks used in our user study with target pose images (first row), skeleton information (second row), and our proposed 3D human shapes in two viewpoints (third row).

$$R = \frac{RMSE_0 - RMSE_{min}}{RMSE_0} \times 100$$

where $RMSE_0$ denotes the initial value of $RMSE$, and $RMSE_{min}$ denotes the minimum value of $RMSE$ at time $t_{min}$. Herein, we use $t_{min}$ as the training time in comparing the training guidance system.

### 6.3 Questionnaire

The questionnaire for comparison study of two training guidance systems was designed as shown in Table 2. There are 8 questions with 5-point Likert scale and two alternatives. The questions were designed to confirm the use of the proposed visual feedback system in the user study from the aspect of different viewpoints, usage of 3D human shape, markers-based guidance, and functions.

All participants in user study were asked to spend around 25 seconds for each training task and answer the questionnaire from his usage experience.

| No | Questions | Answers |
|---|---|---|
| 1 | Have you ever experienced a core training? | YES・NO |
| 2 | Which condition is easier for understanding your pose? | 2D Skeleton・3D Model |
| 3 | Was it easy to understand your pose using two different view windows? | 5・4・3・2・1 |
| 4 | Was it easy to understand your pose by changing the viewpoints? | 5・4・3・2・1 |
| 5 | Was it easy to understand your pose by using the 3D human shape? | 5・4・3・2・1 |
| 6 | Was the marker guidance helpful in correcting your pose? | 5・4・3・2・1 |
| 7 | Did you notice that the sizes of the markers become smaller when they get closer? | YES・NO |
| 7-1 | Was it easy to see the size change of markers? (If Q7's answer is YES) | 5・4・3・2・1 |
| 8 | How much did the estimated pose agree with your understanding? | 5・4・3・2・1 |

Table 2. Questions listed in the questionnaire presented to the users after the user study

## 7. Results

Figure 10 shows a user performing core training using the proposed visual feedback from the training user interface. The results of both objective and subjective evaluations are discussed below.

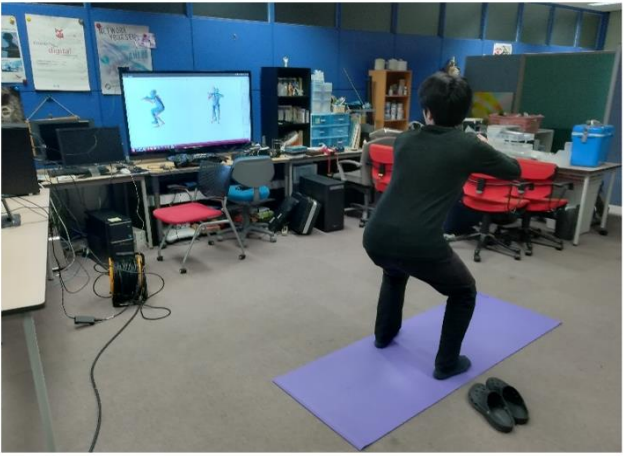

Figure 10. Core training using the proposed visual feedback system.

## 7.1 Objective Evaluation

From the analysis of the results obtained from the user study, we found that the training tasks 3 and 4 can easily be achieved without having large differences between current and target poses. The calculated average training accuracy using skeleton guidance is 33.88, and it becomes 40.18 when the proposed guidance was used. Figure 11 shows the boxplots of both guidance systems. This shows that the proposed system is more effective in correcting postures when performing core training.

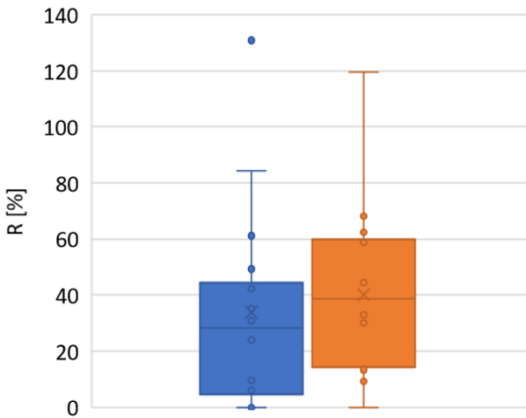

Figure 11. Comparison of training accuracy between skeleton guidance (blue) and the proposed guidance (orange).

Figure 12 shows a comparison of valid training time between skeleton guidance and the proposed system. This result shows that the proposed system requires less time to achieve the correct posture. The conventional guidance system required around 14 seconds to achieve the target pose, while our proposed training system required 12 seconds.

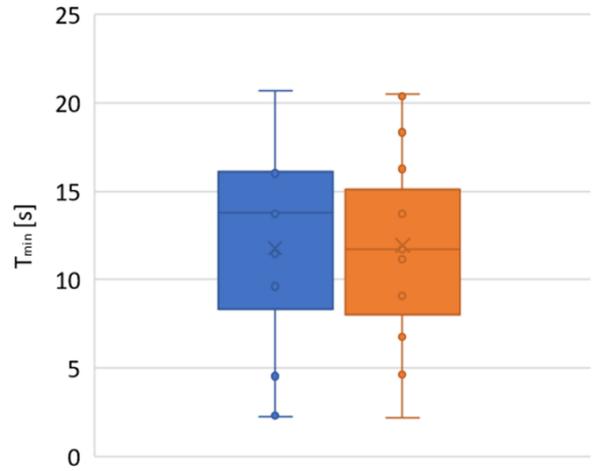

Figure 12. Comparison of training time between skeleton guidance (blue) and the proposed guidance (orange).

## 7.2 Subjective Evaluation

Table 3 shows the results of questionnaires from our user study. All the participants agreed that the proposed guidance using 3D human model is easier for understanding target posture (Question 2). Most of the participants recognized that it becomes easy to understand the target posture with changing viewpoints (Question 4, an average score of 4.38) and 3D human shape (Question 5, an average score of 4.75).

The color-changing markers is thought to be useful in performing core training. However, only one participant noticed that the change in marker sizes in the training may be due to small maker sizes. We observed that the participants preferred checking the whole human mesh in pose correction when they became familiar with the proposed system. We think that color change in the visual feedback on the whole body meshes may have better support in the training process. Additionally, the estimated human model somehow mismatched the user perception (Question 8, an average score of 3.88) because of the limitations of the proposed pose estimation method. This issue could be solved by allowing the user to correct the generated human model interactively in the pre-training user interface (Figure 5).

| Participants | Questions | | | | | | | | |
|---|---|---|---|---|---|---|---|---|---|
| | 1 | 2 | 3 | 4 | 5 | 6 | 7 | 7-1 | 8 |
| A | ○ | 2 | 3 | 4 | 4 | 2 | × | - | 4 |
| B | × | 2 | 3 | 4 | 5 | 2 | × | - | 3 |
| C | ○ | 2 | 2 | 4 | 5 | 4 | × | - | 4 |
| D | ○ | 2 | 5 | 5 | 5 | 5 | ○ | 5 | 5 |
| E | × | 2 | 5 | 5 | 5 | 5 | × | - | 5 |
| F | ○ | 2 | 5 | 3 | 5 | 4 | × | - | 3 |
| G | ○ | 2 | 4 | 5 | 4 | 3 | × | - | 3 |
| H | × | 2 | 3 | 5 | 5 | 3 | × | - | 4 |

Table 3. Questionnaire results. For question 1 and 7, ○ denotes the answer of YES and × denotes the answer of NO; For question 2, 1 denotes 2D skeleton and 2 denotes 3D model.

## 8. Conclusion

In this work, we proposed CoreUI, a visual guidance system for performing core training using a standard web camera. In pose estimation step, we applied deep learning based approaches in 3D human shape and pose estimation by combining OpenPose and human mesh recovery methods. During the runtime process, we estimated the 3D human models from the target image and current camera frame using the proposed pose estimation strategy. Finally, we proposed a user interface for supporting core training with a guide on the differences in markers' positions on the human model. We evaluated the performance of the proposed system through the user study. We conducted the user study by comparing its performance with that of the conventional skeleton-based training guidance method.

As a limitation of this study, the current execution time of the proposed visual feedback system is about 2 seconds for one frame. Although it may be suitable for core training because the user has to maintain the pose over a long period, it will be bottle-neck for other sports such as gymnastics and ball sports where movement is in high speed. We believe this issue can be improved by predicting the training motions with a hallucinatory network [16]. Furthermore, the parallel computing on GPUs with high specifications is preferred. The visual feedback for the high-speed sports can include not only the static pose and also high-level guidance on velocity or momentum. Additionally, the standard SMPL model was used in this study and the actual human body cannot be captured in real-time. It is interesting that we can apply the training guidance on different users considering their body types. In the current implementation, the body shape alignment was based on the root point, it will be helpful to align at the foot joints.

In future work, we plan to use a head-mounted display for the user to have good viewpoints in a virtual environment. We also plan to apply the proposed visual feedback system to other training and sports in the near future.